\documentclass[twocolumn,superscriptaddress]{revtex4}

\usepackage{amssymb}
\usepackage{amsmath}

\usepackage{graphicx}
\usepackage{dcolumn}
\usepackage{bm}

\newfont{\Fr}{eufm10}

\begin{document}

\title{Supernova light-curve fitters and Dark Energy}
\author{Gabriel R. Bengochea}
\email{gabriel@iafe.uba.ar}
\affiliation{Instituto de Astronom\'\i a y F\'\i sica del Espacio
(IAFE), CC 67, Suc. 28, 1428 Buenos Aires, Argentina}

\begin{abstract}

We show that when a procedure is made to remove the tension
between a supernova Ia (SN Ia) data set and observations from BAO
and CMB, there might be the case where the same SN Ia set built
with two different light-curve fitters behaves as two separate and
distinct supernova sets, and the tension found by some authors
between supernova sets actually could be due to tension or
inconsistency between fitters. We also show that the information
of the fitter used in an SN Ia data set could be relevant to
determine whether phantom type models are favored or not when such
a set is combined with the BAO/CMB joint parameter.

\end{abstract}

\pacs{Valid PACS appear here}
 \maketitle

\section{Introduction}

The type Ia supernova (SN Ia) measurements remain a key ingredient
in all current determinations of cosmological parameters. More
than a decade ago, combined observations of nearby and distant SNe
Ia led to the discovery of the accelerating universe picture. It
has become clear that different cosmological observations, such as
the dimming of distant SNe Ia \cite{perlmutter, riess, union2},
anisotropies in the cosmic microwave background (CMB)
\cite{wmap7}, and the signature of baryon acoustic oscillation
(BAO) \cite{eisen, percival} cannot be explained with a
cosmological model that contains only baryonic and dark matter
according to a FRW standard model. The most popular solution is to
introduce an extra component with negative pressure, the so-called
dark energy (e.g., \cite{ht, sahni, padma, frieman, huterer}).

Characterization of dark energy focuses on estimation of the
equation of state $w$, which is the ratio of pressure to density.
For the time-invariant $w=-1$, the equation of state is consistent
with a cosmological constant. Any other fixed or time variable
value of $w$ would require more exotic models.

It is a known fact that the same SN Ia data set in which distance
estimates are analyzed with two different light-curve fitters, the
values achieved for various cosmological parameters (for example,
the equation of state of dark energy) differ, or also there could
be found that some cosmological models result more favored than
others (e.g. \cite{hicken, kessler, sollerman, sanchez, carneiro,
smale}).

Here we analyze the consistency between the fitters MLCS2k2
\cite{mlcs} (hereafter MLCS) and SALT2 \cite{salt2} from a
particular approach as it will be further explained. The
Multicolor Light Curve Shape fitter, MLCS, is the most recent
incarnation of the fitter used by the High-z Supernova Team
\cite{riess}, whilst the Spectral Adaptive Light curve Template,
SALT2, is an improved version of the fitter used originally by the
Supernova Cosmology Project \cite{perlmutter}. A detailed
description of both fitters and a thorough discussion about
systematic errors in SN surveys can be found for example in
\cite{hicken, kessler}.

Each method results in a distance modulus for each supernova.
However, distance moduli calculated for the same objects by the
two fitting methods are not necessarily equal. Whereas the MLCS
calibration uses a nearby training set of SNe Ia assuming a close
to linear Hubble law, SALT2 uses the whole data set to calibrate
empirical light curve parameters. SNe Ia from beyond the range in
which the Hubble law is linear are used, so a cosmological model
must be assumed in this method. Typically a $\Lambda$CDM or a
$w$CDM ($w=const$) model is assumed. Consequently the published
values of SN Ia distance moduli obtained with SALT2 fitter retain
a degree of model dependence. Regarding this, in \cite{frieman1}
it was pointed out that systematic errors in the method of SNe Ia
distance estimation have come into sharper focus as a limiting
factor in SN cosmology. The major systematic concerns for
supernova distance measurements are errors in correcting for
host-galaxy extinction and uncertainties in the intrinsic colors
of supernovae, luminosity evolution, and selection bias in the
low-redshift sample. Also, SALT2 fitter does not provide a
cosmology-independent distance estimate for each supernova, since
some parameters in the calibration process are determined in a
simultaneous fit with cosmological parameters to the Hubble
diagram.

In \cite{wei1} it was investigated the tension between SN Ia data
sets (Constitution \cite{hicken} and Union \cite{kowalski}) and
other data sets, CMB and BAO. There, it was shown that SN Ia data
sets are in tension not only with the observations of CMB and BAO,
but also with other SN Ia data sets such as Davis07
\cite{davis07}. It was also shown that in the Davis07 data set
there is no tension with CMB and BAO observations, concluding that
Union and Constitution data sets are in tension not only with the
observations of CMB and BAO, but also with other SN Ia data sets.
Then, the author found the main sources responsible for the
tension by employing a truncation method, following the simple
procedure used in \cite{nesseris}. With this in mind, the
truncated UnionT and ConstitutionT data sets were built, which are
consistent with the other observations and the tension was
completely removed.

In the quest of the characterization of dark energy, in
\cite{cal02} it was first noted that observational data do not
rule out the possibility that $w<-1$. Phantom dark energy models
with $w<-1$ have the interesting properties that the density of
the dark energy increases with increasing scale factor, and the
phantom energy density can become infinite at a finite time, a
condition known as the 'big rip'.

Future projects as SNAP \cite{snap} are designed to reveal the
nature of the dark energy. It will characterize the dark energy
density, equation of state and time variation by precisely and
accurately measuring the distance-redshift relation of SNe Ia. The
matter density, dark energy density, and flatness of the universe
could be determined at the 1\% level, including systematic
uncertainties, the dark energy equation of state to about 3\% and
its time variation characterized to within 10\% of the Hubble
expansion time. In summary, this kind of project will seek to
determine whether $w=-1$ or not, and if $w$ is a constant or it
evolves in time. However, in order to do this, the matter of
reducing the systematic uncertainties between light-curve fitters
will be of great importance. The combination of observational data
sets might lead, as it will be further shown, the equation of
state of dark energy to be of the phantom type or not, depending
on how the SN Ia data were processed.

In this Letter we focus on the consistency of two of the main
light-curve fitters used for the elaboration of SN Ia data sets.
To accomplish this, we present another approach to the tension
between SN Ia data sets as the one found in \cite{wei1} and then
the consistency between light-curve fitters is analyzed when
appears the need of removing tension between an SN Ia data set and
BAO/CMB data applying a truncation method to the same SN Ia data
set, but obtained with the two different fitters. Additionally, we
show that the conclusion about if the combination of SN Ia data
with BAO/CMB favors or not an equation of state for dark energy of
the phantom type in the framework of a given cosmological model
might depend, in some cases, on the fitter employed in the
elaboration of the SN Ia set used in the analysis.

\section{Tension between supernovae Ia data sets revisited}

In this section and the next one we will use a
$\chi^2=\chi^2_{SNe}+\chi^2_{BAO/CMB}$ statistic to analyze the
confidence intervals of the free parameters of two cosmological
models, by employing different SN Ia data sets and their
combination with the BAO/CMB joint parameter introduced in
\cite{sollerman}. Since lately more non-standard models are built
to try to explain the dark energy phenomenon, we chose to combine
the SNe analysis with the BAO/CMB joint parameter which is better
suited for these sort of models, using it only as an example of
data combination. We must also clarify that the goal of this work
was neither to put constraints nor to find best fits to
cosmological models, but to show certain extra information that
should be minded when non-SNe data sets are combined with SNe data
sets analyzed with different fitters. The separate $\chi^2$ of SNe
Ia and BAO/CMB used in this work are shown in Appendix A.

We considered for the analysis the $\Lambda$CDM and flat $w$CDM
models with $w=const$. In the case of $\Lambda$CDM the dark energy
is a cosmological constant which behaves as a vacuum energy with
$w=-1$, but we allowed non-zero spatial curvature $\Omega_k$. Then
the dimensionless expansion rate is given by,
\begin{eqnarray} \nonumber
&&E\equiv H(z)/H_0=\\
&&[\Omega_m(1+z)^3+\Omega_r(1+z)^4+\Omega_k(1+z)^2+\Omega_{\Lambda}]^{1/2}\:\:\:\:\:\:\:\:
\label{HLCDM}
\end{eqnarray}
where $H(z)$ is the Hubble parameter as a function of the
cosmological redshift $z$; $\Omega_m$, $\Omega_r$ and
$\Omega_{\Lambda}$ are the contributions of matter, radiation and
dark energy respectively to the total energy density today, and
the curvature density is
$\Omega_k=1-\Omega_m-\Omega_r-\Omega_{\Lambda}$. The parameters
usually chosen as free parameters in this model are $\Omega_m$ and
$\Omega_{\Lambda}$.

For the flat $w$CDM ($\Omega_k=0$) case, we allowed the equation
of state parameter of dark energy $w$ to differ from -1 so,
\begin{equation}
E=[\Omega_m(1+z)^3+
\Omega_r(1+z)^4+\Omega_{\Lambda}(1+z)^{3(1+w)}]^{1/2}
\label{HwCDM}
\end{equation}
where now $\Omega_{\Lambda}=1-\Omega_m-\Omega_r$. The free
parameters in this model are $\Omega_m$ and $w$. The case flat
$\Lambda$CDM is a special case of this one.

The choice of the considered models for this analysis was
arbitrary and not relevant, this was for mere simplicity (two free
parameters) with the aim of showing the novel results. The
consideration of models with an equation of state $w$ that does
not evolve in time is enough, since the current data do not yield
precise constraints on the time derivative of $w$. (For
constraints on time-varying $w$, see \cite{sollerman} as an
example). Additionally, the rationale for the flat case of $w$CDM
was that the WMAP data from the CMB anisotropy constrain the
spatial curvature to be very small (e.g. \cite{komatsu}). Also,
the use of these two models is the most suitable to obtain a
correct comparison among different fitters, since the light-curve
fitter SALT2 uses the whole data set to calibrate empirical light
curve parameters and a $\Lambda$CDM or a $w$CDM model is typically
assumed, as it was mentioned in Section I.

As it was previously mentioned, the analysis in this work was
performed in the framework of SALT2 \cite{salt2} and MLCS
\cite{mlcs} fitters and the SN Ia data sets used were Constitution
data sets (Tables 2 and 4 from \cite{hicken} as it will be further
explained in Section III), the SDSSII full data set (Tables 10 and
14 from \cite{kessler} with the same values used for the
'intrinsic' dispersions there) and the Union2 data set
\cite{union2}.

It was interesting to observe the discrepancy between the results
obtained when using one fitter or the other when it is allowed the
variation of $\Omega_k$ in the framework of a $\Lambda$CDM
($w=-1$) model. The analysis of the SDSSII full data set with 288
SNe Ia built with MLCS showed that the flat case ($\Omega_m=0.27$
\cite{wmap7}) stands excluded to more than 3 $\sigma$ confidence
level, while with the same data set, but processed with SALT2 this
did not happen at all (Figs. 1a and 1c). Something similar happens
in the framework of the flat $w$CDM model (Figs. 1b and 1d): the
standard model ($\Omega_m=0.27, w=-1$) using SDSSII (MLCS) is
excluded to more than 2 $\sigma$ confidence level (see also Fig.1
of \cite{sollerman}). Since the responsible of this fact is the
fitter and not the SNe Ia (because the data set is the very same
and only the fitter was changed), one could then wonder what SN Ia
data set should be used to be combined with, for example CMB data,
which leave little margin to the variation of $\Omega_k$ (e.g.
\cite{komatsu}). Looking at Figs. 1c and 1d one would choose those
data sets processed with SALT2; however we should keep in mind
that SALT2 fitter retains a degree of model dependence because
typically a $\Lambda$CDM model is assumed.
\begin{figure}[h!]
\begin{center}
\includegraphics[width=9cm,angle=0]{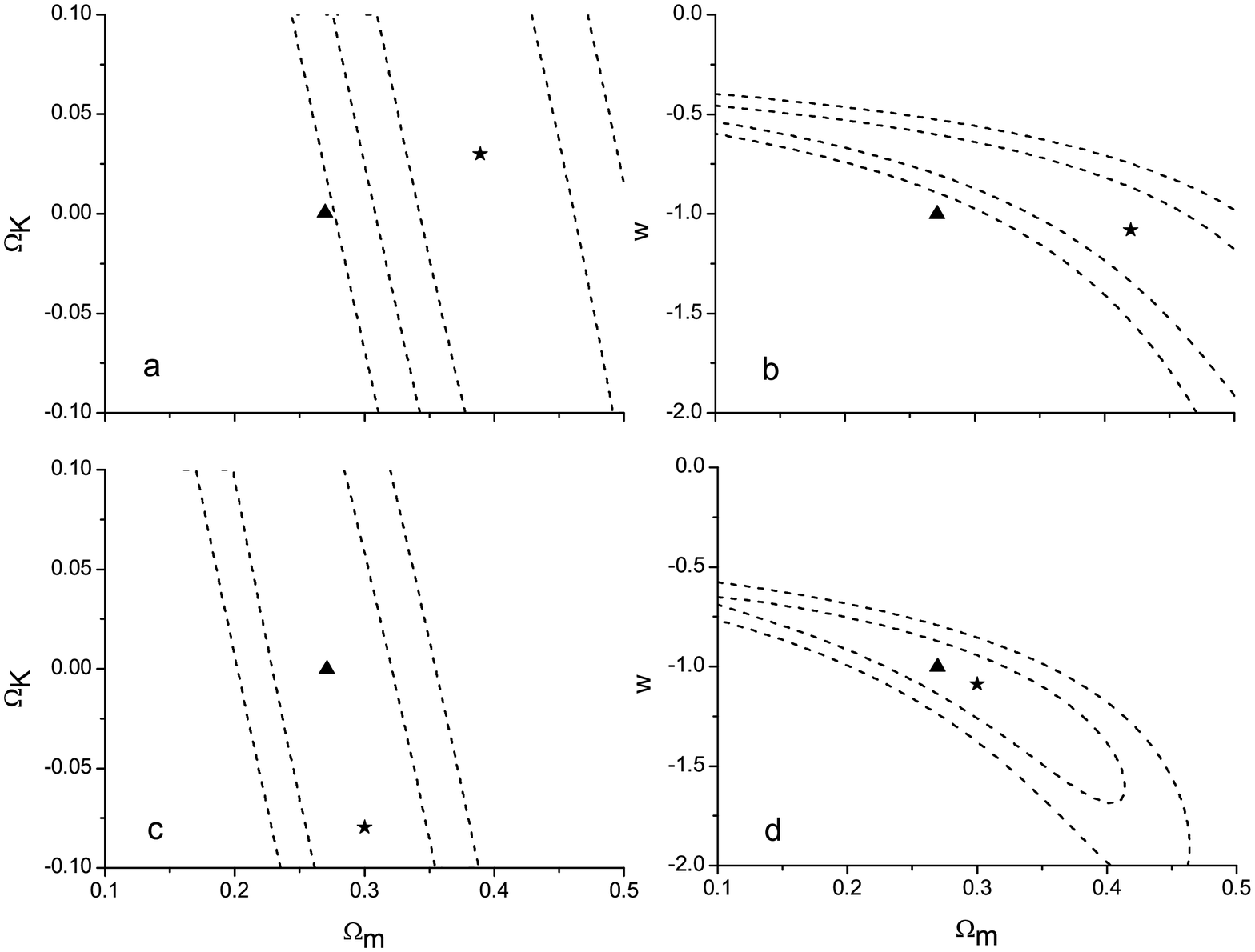}
\end{center}
\caption{ (a) Confidence intervals at 68.3\%, 95.4\% and 99.7\% in
the $\Omega_m-\Omega_k$ plane for the SDSSII (MLCS) SN Ia data set
in the $\Lambda$CDM model framework (dashed lines). (b) Confidence
intervals at 68.3\% and 95.4\% in the $\Omega_m-w$ plane for the
SDSSII (MLCS) SN Ia data set in the flat $w$CDM model framework
(dashed lines). (c) Confidence intervals at 68.3\% and 95.4\% in
the $\Omega_m-\Omega_k$ plane for the SDSSII (SALT2) SN Ia data
set in the $\Lambda$CDM model framework (dashed lines). (d)
Confidence intervals at 68.3\% and 95.4\% in the $\Omega_m-w$
plane for the SDSSII (SALT2) SN Ia data set in the flat $w$CDM
model framework (dashed lines). The best fits are indicated with a
star whereas the standard flat $\Lambda$CDM ($\Omega_m=0.27$) is
marked with a triangle. } \label{Fig1}
\end{figure}

In \cite{wei1}, it was found that Union \cite{kowalski} and
Constitution \cite{hicken} data sets were in tension not only with
BAO and CMB data, but with other SN Ia data sets too (e.g. Davis07
\cite{davis07}). The tension found was attributed to certain
supernovae of the data set and by a truncation method these
outliers were removed from the set with the objective of releasing
the tension.

We found interesting to analyze what would happen if we applied a
criterion in order to study the consistency between data sets, a
criterion more restrictive than the only fact that the confidence
intervals overlap. To perform this analysis, we adopted the
criterion of considering the existence of tension between a given
data set and another set constituted combining several data sets
(including the first one) as the fact that the best fit point to
the first data set is out of the 68.3\% (1$\sigma$) confidence
level contour given by the combined data set. Similar criteria
were adopted in their analysis by \cite{nesseris, wei1, li, gb}.
This is a criterion we will adopt in order to show how differently
will behave the same SN Ia set processed with two different
fitters when a truncation method is performed. This will lead to
an alternative way of analyzing the discrepancy between the
results obtained when one or other fitter is used. One could
choose not to use this more restrictive criterion, nevertheless
with this adopted criterion, we seek more physical consistency
between best-fits, so the best fits do not drive to too different
cosmological evolutions. A best fit which effective equation of
state is of the phantom type \cite{cal02} ($w<-1$) tells us about
very different physics from the one that is not. For instance, in
a recent work \cite{gb} the consequences of applying it to several
data sets in the framework of $f(T)$ theories have been
investigated.

When we used the Union2 recently released data set with 557 SNe Ia
(processed with SALT2 fitter) to combine it with the information
from BAO/CMB in the framework of the flat $w$CDM model, we found
that there is no tension between data sets (Fig. 2a). However,
when we did the same procedure with the 288 SNe Ia of the SDSSII
set (in the framework of MLCS fitter) we found that there is
tension between SNe Ia and BAO/CMB to more than 3 $\sigma$ (Fig.
2b). Since the data from BAO/CMB are the same used in both cases,
one could wrongly conclude that there is tension between the SN Ia
data sets. This case is similar to what was found in \cite{wei1}.
\begin{figure}[h!]
\begin{center}
\includegraphics[width=8cm,angle=0]{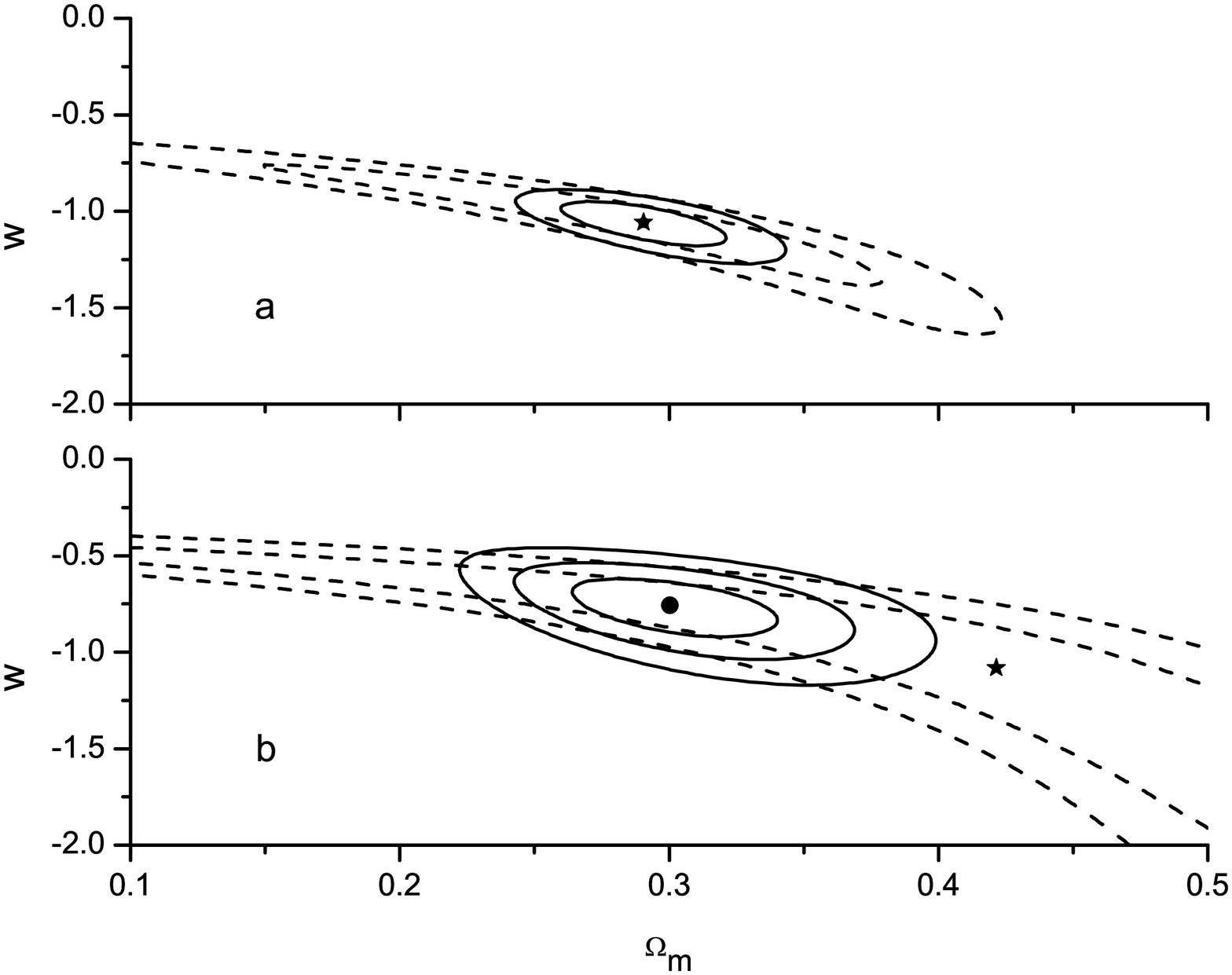}
\end{center}
\caption{(a) Confidence intervals at 68.3\% and 95.4\% in the
$\Omega_m-w$ plane for the Union2 (SALT2) SNe Ia only (dashed
lines) and Union2+BAO/CMB (solid lines). (b) Confidence intervals
at 68.3\%, 95.4\% and 99.7\% in the $\Omega_m-w$ plane for the
SDSSII (MLCS) SNe Ia only (dashed lines) and SDSSII+BAO/CMB (solid
lines). The best fits to SNe are indicated with a star whereas the
best fits to the combined parameters are indicated with a dot. }
\label{Fig2}
\end{figure}
The result obtained in \cite{wei1}, where there is tension between
Union and BAO, while in Davis07 there is none, seems to be due to
a data set (Union) is processed with SALT and the other (Davis07)
with versions of MLCS. The tension between Union and Davis07 could
actually be a trouble between fitters and not between supernovae
sets as it will be discussed in the next section.

To better understand the analysis of tension between Union2 and
SDSSII performed here, in Figs. 3a and 3b are displayed the
combinations of SNe Ia+BAO/CMB for the Union2 and SDSSII cases
respectively. For the case Union2 vs SDSSII(MLCS) there is tension
to more than 2 $\sigma$ level, while in Union2 vs SDSSII(SALT2)
there is none. Clearly the tension exists between fitters and not
between SN Ia data sets.

\begin{figure}[h!]
\begin{center}
\includegraphics[width=8cm,angle=0]{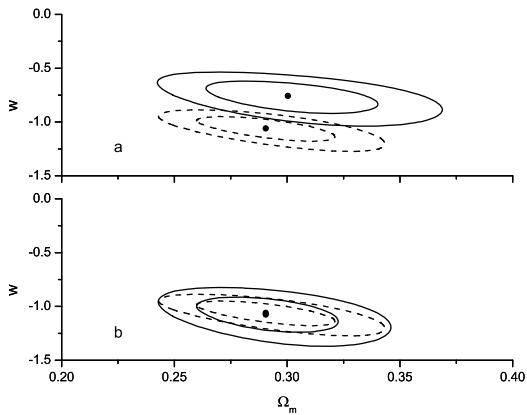}
\end{center}
\caption{Confidence intervals at 68.3\% and 95.4\% in the
$\Omega_m-w$ plane from combining SNe Ia+BAO/CMB. (a) Union2
(SALT2 - dashed lines) vs SDSSII (MLCS - solid lines). (b) Union2
(SALT2- dashed lines) vs SDSSII (SALT2 - solid lines). A tension
between fitters is revealed. } \label{Fig3}
\end{figure}

\section{Are supernovae light-curve fitters consistent?}

In this section we analyze the problem of tension between
light-curve fitters and the consistency between them from a
different approach.

We found interesting to study what happens when one needs to
perform a truncation procedure as in \cite{wei1} to remove the
tension between BAO/CMB and an SN Ia data set (processed with a
given fitter) and the same SN Ia data set but with a different
fitter. It is important to emphasize that the adopted criterion
itself is not relevant, but how the same SN set behaves when it is
processed with different fitters.

The criterion consists in finding and removing the outliers
responsible of the tension. First, we fitted the model to the
whole SN Ia data set finding the best fit parameters, including
the nuisance parameter $\mu_0$ (see Appendix A). Then, we
calculated the relative deviation to the best fit prediction,
$\mid\mu_{obs}-\mu_{th}\mid/\sigma_{obs}$, for all the data points
finding which cut solved the tension problem and which SNe Ia were
the outliers.

We performed this procedure in the framework of the flat $w$CDM
model for two different SN Ia data sets, Constitution (SALT2 and
MLCS) and SDSSII (SALT2 and MLCS).

For the data sets here denominated Constitution (SALT2) and
Constitution (MLCS) we used the same 337 SNe Ia from the Table 2
(SALT2) and Table 4 (MLCS17) from \cite{hicken}. In \cite{hicken}
MLCS was used to find the dust-reddening properties through the
value of $R_V$ that minimizes the scatter in the Hubble residuals
for the nearby CfA3 sample and they found $R_V=1.7$. MLCS with
$R_V=3.1$ overestimates host-galaxy extinction while $R_V=1.7$
does not. With its lower value of $R_V$, MLCS17 attributes less
host extinction to each SN Ia and therefore this produces a larger
distance compared to MLCS31.

Figure 4a shows the confidence intervals to 68.3\% and 95.4\% in
the plane $\Omega_m-w$ for the Constitution (SALT2) data set of
SNe Ia only and for the combination SNe Ia+BAO/CMB. There, a
tension to 2 $\sigma$ confidence level between both data sets can
be observed. In Fig. 4b it is shown that after the truncation
procedure with a 2.1 $\sigma$ cut (13 outliers) the tension was
removed completely. In a similar way, we proceeded to do the same
with the Constitution (MLCS) data set and the results are shown in
Figs. 5a and 5b. In this case, a 2 $\sigma$ cut was enough and 12
SNe Ia were removed from the data set so the tension with BAO/CMB
was removed. Although the necessary cut for Constitution set was
slightly different for SALT2 than for MLCS, and although the
number of SNe to remove was not significantly different, the most
remarkable fact was that \emph{only 4 outliers were the same}.
This can be clearly seen in Table \ref{FigT} where SNe Ia outliers
are displayed for each case and which of those are the ones in
common. We want to stress that the SN Ia data set was the same and
the only difference lied in the light-curve fitter employed. Then,
with this truncation procedure we found \emph{the behavior was as
if there were two different SN Ia sets when actually there was
only one}.

\begin{figure}[h!]
\begin{center}
\includegraphics[width=8cm,angle=0]{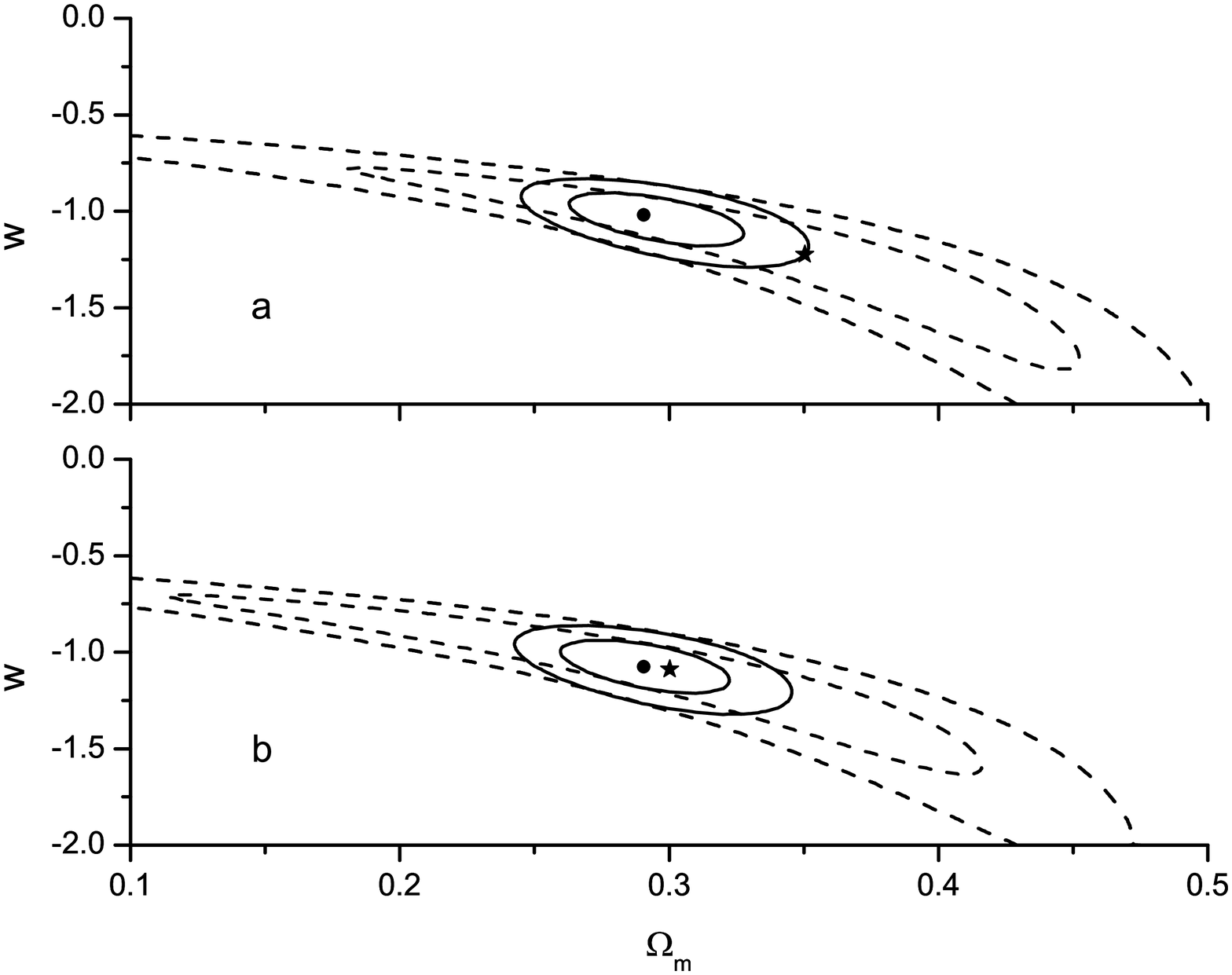}
\end{center}
\caption{(a) Confidence intervals at 68.3\% and 95.4\% in the
$\Omega_m-w$ plane for the Constitution (SALT2) SNe Ia only
(dashed lines) and SNe+BAO/CMB (solid lines). Tension can be
appreciated between data sets. (b) Idem (a), but after performing
the truncation procedure. The best fits to SNe are indicated with
a star whereas the best fits to the combined parameters are
indicated with a dot.} \label{Fig4}
\end{figure}

\begin{figure}[h!]
\begin{center}
\includegraphics[width=8cm,angle=0]{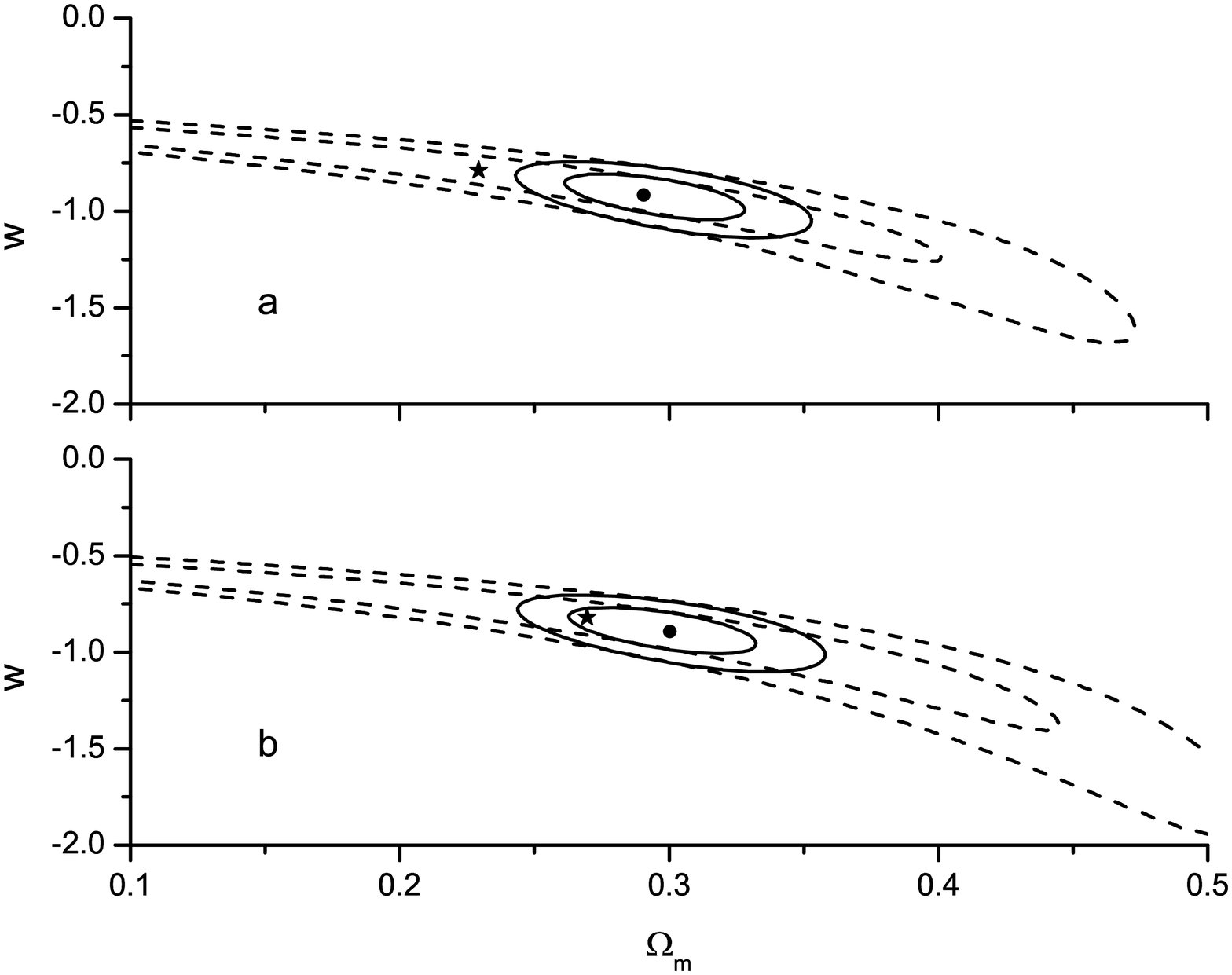}
\end{center}
\caption{(a) Confidence intervals at 68.3\% and 95.4\% in the
$\Omega_m-w$ plane for the Constitution (MLCS) SNe Ia only (dashed
lines) and SNe+BAO/CMB (solid lines). Tension can be appreciated
between data sets. (b) Idem (a), but after performing the
truncation procedure. The best fits to SNe are indicated with a
star whereas the best fits to the combined parameters are
indicated with a dot.} \label{Fig5}
\end{figure}

\begin{table}[h!]
\caption{ The outliers after the truncation procedure by using the
same Constitution SN Ia data set processed with MLCS and SALT2
light-curve fitters. Note that there are only 4 matches in the
outlier SNe.}
\begin{center}
\includegraphics[width=4cm,angle=0]{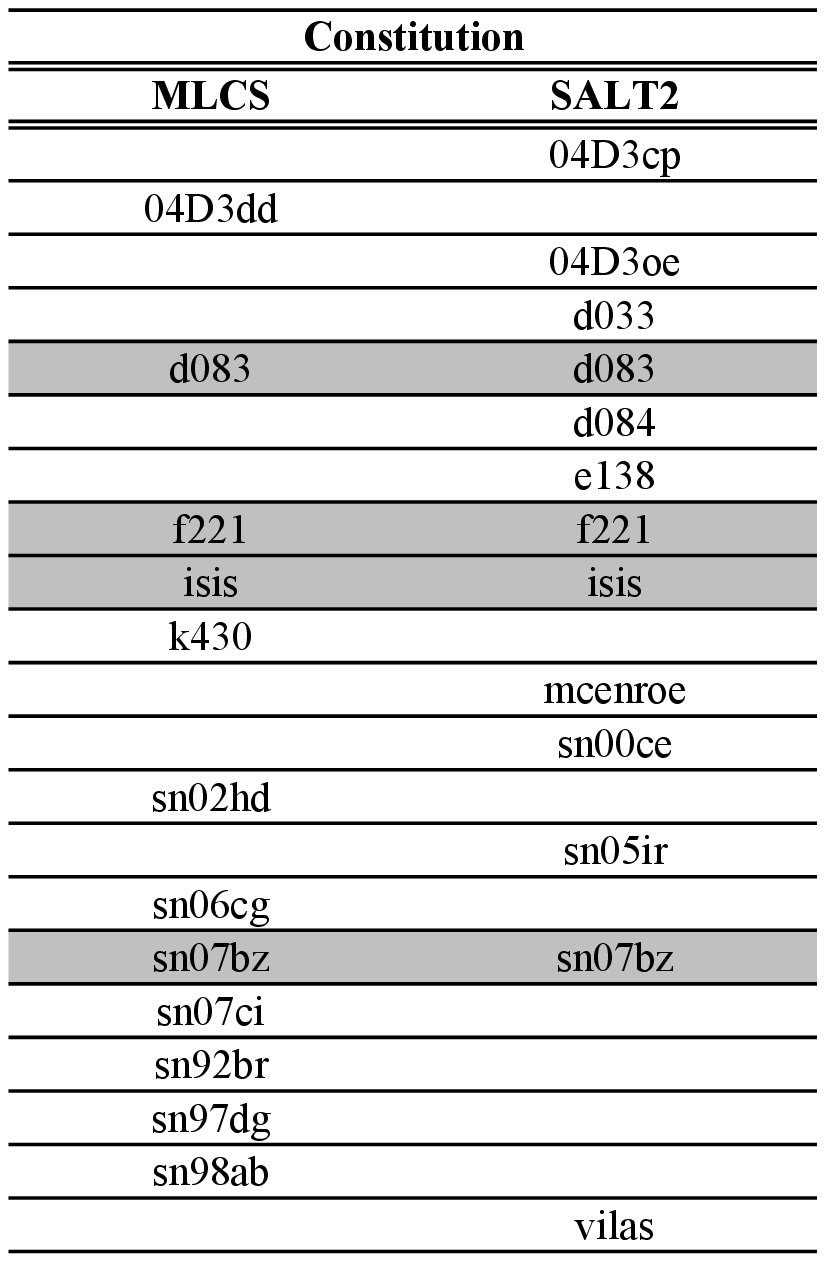}
\end{center}
\label{FigT}
\end{table}

Something more drastic occurred when we did the same analysis, but
using the SDSSII data set. When the SDSSII (SALT2) data set was
used, no SNe Ia needed to be removed, since there was no tension
with BAO/CMB (Fig. 6a). However, with the SDSSII (MLCS) data set
there was tension with BAO/CMB to a level greater than 99.7\%
(Fig. 6b) and a cut of 1.7 $\sigma$ (18 outliers) was needed to
remove such tension. To highlight this, in Fig. 7 we show the
confidence intervals to 99.7\% of the combined SNe Ia+BAO/CMB for
the cases SDSSII (SALT2) and SDSSII (MLCS). There, it can be seen
how both best fits differ by more than 3 $\sigma$ level. Another
interesting thing is the comparison of Figs 7 and 3a. This
comparison allows appreciating how \emph{two light-curve fitters
employed for the same SN Ia set produce the same result than two
different SN Ia sets}.
\begin{figure}[h!]
\begin{center}
\includegraphics[width=8cm,angle=0]{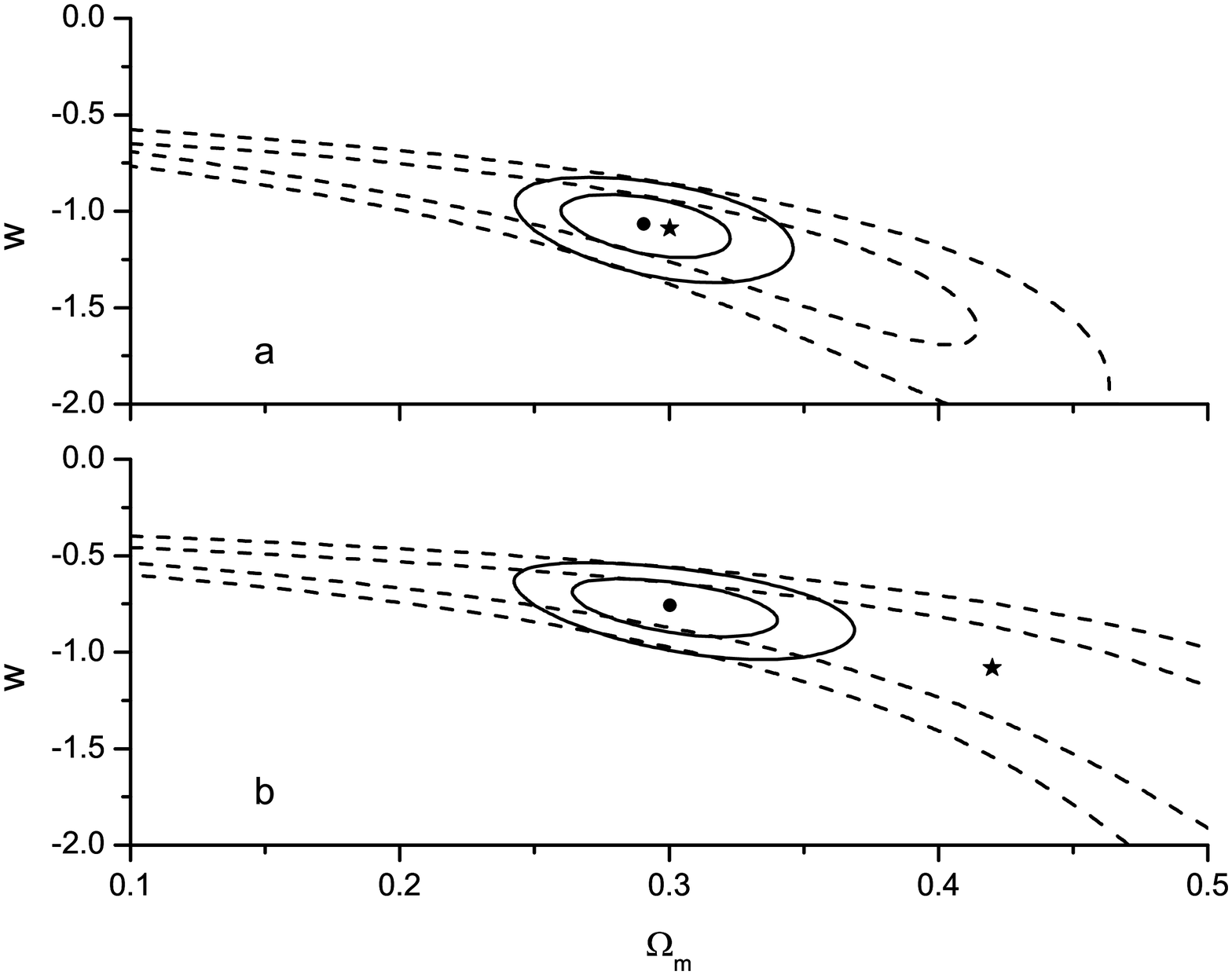}
\end{center}
\caption{(a) Confidence intervals at 68.3\% and 95.4\% in the
$\Omega_m-w$ plane for the SDSSII (SALT2) SNe Ia only (dashed
lines) and SNe+BAO/CMB (solid lines). There is no tension between
the data sets. (b) Idem (a), but for the SDSSII (MLCS) SNe Ia.
Here can be appreciated tension to more than 2 $\sigma$ confidence
level.} \label{Fig6}
\end{figure}

\begin{figure}[h!]
\begin{center}
\includegraphics[width=7cm,angle=0]{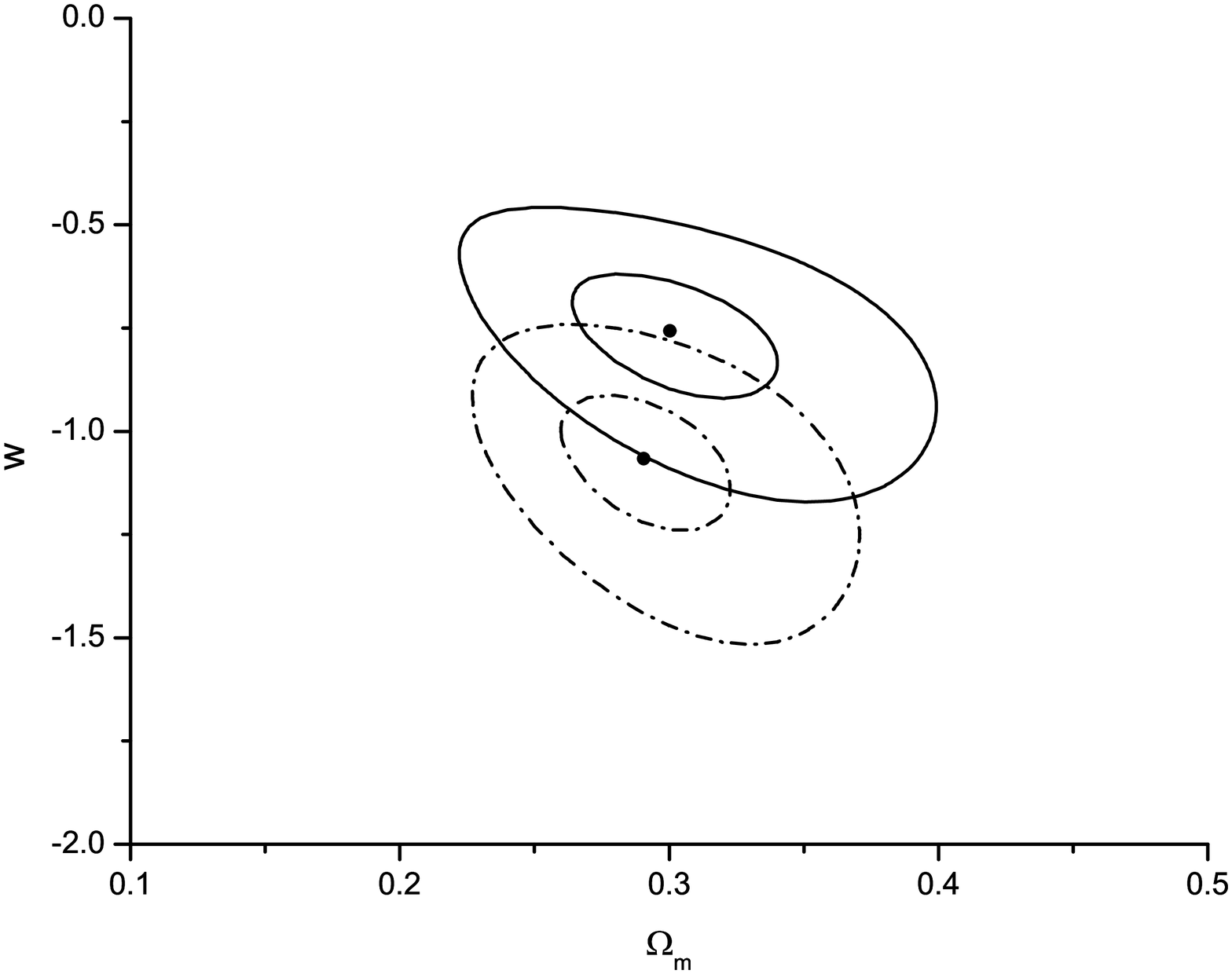}
\end{center}
\caption{Confidence intervals at 68.3\% and 99.7\% in the
$\Omega_m-w$ plane coming from combining SN Ia and BAO/CMB data.
Solid lines correspond to SDSSII (SALT2) whereas dash-dotted lines
correspond to SDSSII (MLCS). There can be seen that both best fits
are outside 3 $\sigma$ confidence level from the other fitter.}
\label{Fig7}
\end{figure}

The fact of needing to take out different supernovae to remove the
tension in Constitution (SALT2) and in Constitution (MLCS) reveals
that they are not the supernovae themselves that probably generate
the tension, but the fitters. Using the same SN Ia data set with
the same truncation process, the SNe Ia we had to take out were
not the same ones. It seems as if a given fitter makes some
supernovae bring an apparent problem. The case of the SDSS data
set also presented this behavior, because with one fitter there
were supernovae that caused tension and with the other there were
none, since there was no need of truncation. What we found here
with two different fitters for the same SN Ia data set is analogue
to what was found in \cite{wei1} between two different SN Ia sets.
What was found between Union and Davis07 data sets is perhaps a
fitter problem and not supernovae.

We also analyzed another interesting aspect. We found important to
stress how an employed fitter in an SN Ia data set could take part
in the conclusion about the equation of state $w$ of a given model
when this SN Ia set is combined with BAO/CMB, being this joint
parameter more suitable for the lately more common non-standard
models. In Table \ref{phantom} (Phantom Table) we show how for two
given theories using the combination BAO/CMB+SN Ia (SALT2) is
obtained a best fit for $w$ of the phantom type, while with MLCS
the opposite occurs. There, the result is shown for the flat
$w$CDM model and for a non-standard model of modified gravity
$f(T)$ as in \cite{tortion, gb} when Union2 (SALT2), SDSSII (SALT2
and MLCS) and Constitution (SALT2 and MLCS) are used.

\begin{table}[h!]
\caption{\emph{Phantom Table}. Best fit $w$ types for two
cosmological models when the BAO/CMB joint parameter is combined
with an SN Ia data set processed with a given light-curve fitter.}

\centering
\begin{tabular}{|c|c|c|} \hline\hline
\bf Model &\bf BAO/CMB\:+ & \bf Best fit $w$ type \\
\hline Flat $w$CDM & Union2 (SALT2)  & Phantom \\
Flat $w$CDM & SDSSII (SALT2) & Phantom \\
Flat $w$CDM & SDSSII (MLCS) & non-Phantom \\
Flat $w$CDM & Const.(SALT2) & Phantom \\
Flat $w$CDM & Const.(MLCS) & non-Phantom\\
$f(T)$ & Union2 (SALT2)& Phantom\\
$f(T)$ & SDSSII (SALT2) & Phantom\\
$f(T)$ & SDSSII (MLCS) & non-Phantom\\
$f(T)$ & Const.(SALT2) & Phantom\\
$f(T)$ & Const.(MLCS) & non-Phantom\\
\hline\hline
\end{tabular}
\label{phantom}
\end{table}

We do not mean that this is going to occur with any model, but to
conclude that the combination of data sets favors or not phantom
type models with $w<-1$, the fitter used to process the SNe Ia is
an additional factor that must be taken into account as source of
degeneration.

\section{Conclusions}

Although the evidence in the FRW framework that the universe is
going through an accelerated stage, because of the existence of
what we call dark energy, is solid from various observational data
sets, its nature will depend on the future better understanding of
the systematic errors present in supernovae observations,
particularly those present in the light-curve fitters analyzed
here.

As it is well known, when the same SN Ia data set is processed
with two different fitters, the values found for cosmological
parameters (such as the equation of state of dark energy) differ.

Here we analyzed this difference showing how when a procedure to
remove tension between an SN Ia data set and observations from
BAO/CMB is performed, there could exist the case where the same SN
Ia set processed with two different fitters behaves as if there
were two different sets, and the tension between sets found by
some authors actually could be due to a tension or inconsistency
between fitters.

We also showed that the information of the fitter used in an SN Ia
set could be relevant and it should be minded as an additional
factor to decide if phantom type models are favored or not when
the given SN Ia set is combined with the BAO/CMB joint parameter.

\bigskip

\acknowledgments{G.R.B. is supported by CONICET. I would like to
thank Tamara Davis and Jesper Sollerman for kind and helpful
discussions about SDSS SNe Ia data sets. Also, I thank Diego
Travieso for his numerical collaboration and interesting
discussions and Rafael Ferraro for his support.}

\appendix

\section{Cosmological constraints methods}

\subsection{Type Ia Supernovae constraints}

The N data points of the SNe Ia compiled in a data set are usually
given in terms of the distance modulus $\mu_{obs}(z_i)$. On the
other hand, the theoretical distance modulus is defined as
\begin{equation}\label{muth}
\mu_{th}(z_i)=5 log_{10} D_L(z_i)+\mu_0
\end{equation}
where $\mu_0\equiv 42.38 - 5 log_{10}h$ and $h$ is the Hubble
constant $H_0$ in units of 100 km/s/Mpc, whereas the Hubble-free
luminosity distance for the general case is,
\begin{equation}\label{dlHfree}
D_L(z)= (1+z)\:|\Omega_k|^{-1/2}\:{\cal S}_k
\:\Big[|\Omega_k|^{1/2}\int_{0}^{z}\frac{dz^{\prime}}{E(z^{\prime},\mathbf{p})}\Big]
\end{equation}
in which $E\equiv H(z)/H_0$ is the dimensionless expansion rate,
$\mathbf{p}$ denotes the model parameters, and the function ${\cal
S}_k(x)=sin(x)$ when the curvature density $\Omega_k<0$, ${\cal
S}_k(x)=sinh(x)$ for $\Omega_k>0$ and ${\cal S}_k(x)=x$ for the
flat case $\Omega_k=0$. Correspondingly, the $\chi^2$ from the N
SNe Ia is given by
\begin{equation}  \label{chi}
\chi^2_{SNe}(\mathbf{p})=\sum_{i=1}^{N}\frac{[\mu_{obs}(z_i)-\mu_{th}(z_i;\mathbf{p})]^2}{\sigma^2(z_i)}
\end{equation}
where $\sigma(z_i)$ is the corresponding uncertainty for each
observed value. The parameter $\mu_0$ is a nuisance parameter but
it is independent of the data points. One can perform an uniform
marginalization over $\mu_0$. However, there is an alternative
way. Following \cite{chi2sn}, the minimization with respect to
$\mu_0$ can be made by expanding the $\chi^2_{SNe}$ of (\ref{chi})
with respect to $\mu_0$ as
\begin{equation}  \label{chib}
\chi^2_{SNe}(\mathbf{p})=\tilde{A}-2\mu_0 \tilde{B}+\mu_0^2
\tilde{C}
\end{equation}
where,
\begin{eqnarray}  \nonumber
\tilde{A}(\mathbf{p})&=& \sum_{i=1}^{N}\frac{[\mu_{obs}(z_i)-\mu_{th}(z_i;\mu_0=0,\mathbf{p})]^2}{\sigma^2(z_i)}\\
\tilde{B}(\mathbf{p})&=& \sum_{i=1}^{N}\frac{[\mu_{obs}(z_i)-\mu_{th}(z_i;\mu_0=0,\mathbf{p})]}{\sigma^2(z_i)}\nonumber\\
\tilde{C}&=& \sum_{i=1}^{N}\frac{1}{\sigma^2(z_i)} \nonumber
\end{eqnarray}
Eq. (\ref{chib}) has a minimum for $\mu_0=\tilde{B}/\tilde{C}$ at
\begin{equation}  \label{chi2posta}
\tilde{\chi}^2_{SNe}(\mathbf{p})=\tilde{A}(\mathbf{p})-\frac{\tilde{B}(\mathbf{p})^2}{\tilde{C}}
\end{equation}
Since $\chi^2_{SNe,min}=\tilde{\chi}^2_{SNe,min}$ obviously, we
can instead minimize $\tilde{\chi}^2_{SNe}$ which is independent
of $\mu_0$.

\subsection{Combined BAO/CMB parameter constraints}

Since in this work we combined CMB and BAO observations with SN Ia
data sets for flat models, we consider here all the relations for
the spatially flat case.

A more model-independent constraint can be achieved by multiplying
the BAO measurement of $r_s(z_d)/D_V(z)$ with the position of the
first CMB power spectrum peak \cite{komatsu} $\ell_A=\pi
d_A(z_*)/r_s(z_*)$, thus canceling some of the dependence on the
sound horizon scale \cite{sollerman}. Here, $d_A(z_*)$ is the
comoving angular-diameter distance to recombination, $r_s$ is the
comoving sound horizon at photon decoupling, $z_d\thickapprox1020$
is the redshift of the drag epoch at which the acoustic
oscillations are frozen in, and $D_V$ is defined as (assumed a
$\Lambda $CDM model) \cite{eisen},
\begin{equation}  \label{DV}
D_V(z)=\Bigg[\frac{z}{H(z)}\:\Big(\int_{0}^{z}\frac{dz^{\prime}}{%
H(z^{\prime})}\Big)^{2}\Bigg]^{1/3}
\end{equation}
We further assume $z_*=1090$ from \cite{komatsu} (variations
within the uncertainties about this value do not give significant
differences in the results).

In \cite{percival} was measured $r_s(z_d)/D_V(z)$ at two
redshifts, $z=0.2$ and $z=0.35$, finding
$r_s(z_d)/D_V(0.2)=0.1905\pm0.0061$ and
$r_s(z_d)/D_V(0.35)=0.1097\pm0.0036$. Combining this with $\ell_A$
gives the combined BAO/CMB constraints \cite{sollerman}:
\begin{eqnarray}\nonumber
\frac{d_A(z_*)}{D_V(0.2)}\:\frac{r_s(z_d)}{r_s(z_*)}&=& 18.32\pm0.59\\
\frac{d_A(z_*)}{D_V(0.35)}\:\frac{r_s(z_d)}{r_s(z_*)}&=&
10.55\pm0.35\label{soll}
\end{eqnarray}

Before matching to cosmological models we also need to implement
the correction for the difference between the sound horizon at the
end of the drag epoch and the sound horizon at last scattering.
The first is relevant for the BAO, the second for the CMB, and
$r_s(z_d)/r_s(z_*)=1.044\pm0.019$ (using values from
\cite{komatsu}). Inserting this into (\ref{soll}) and taking into
account the correlation between these measurements using the
correlation coefficient of 0.337 calculated by \cite{percival},
gives the final constraints we use for the cosmology analysis
\cite{sollerman}:
\begin{eqnarray}  \nonumber
A_1=\frac{d_A(z_*)}{D_V(0.2)}&=& 17.55\pm0.65\\
A_2=\frac{d_A(z_*)}{D_V(0.35)}&=& 10.10\pm0.38\label{bao/cmb}
\end{eqnarray}

Using this BAO/CMB parameter cancels out some of the dependence on
the sound horizon size at last scattering. This thereby removes
the dependence on much of the complex pre-recombination physics
that is needed to determine that horizon scale \cite{sollerman}.
In all the cases, we have considered a radiation component
$\Omega_r$=5 x $10^{-5}$.

So, for our analysis we add to the $\chi^2$ statistic:
\begin{equation}  \label{chibaocmb}
\chi^2_{BAO/CMB}(\mathbf{p})=\sum_{i=1}^{N=2}\frac{[A_{obs}(z_i)-A_{th}(z_i;\mathbf{p})]^2}{\sigma_A^2(z_i)}
\end{equation}
where $\mathbf{p}$ are the free parameters, $A_{obs}$ is the
observed value ($A_1$ and $A_2$), $A_{th}$ is the predicted value
by the model and $\sigma_A$ is the $1\sigma$ error of each
measurement.

\end{document}